# Customizable Laguerre–Gaussian Perfect Vortex Beams

Wenxiang Yan,[1] Zheng Yuan,[1] Yuan Gao,[1] Zhi-Cheng Ren,[1,2] Xi-Lin Wang,[1,2] Jianping Ding,[1,2,3,] and Hui-Tian Wang,[1,2]

[1]National Laboratory of Solid State Microstructures and School of Physics, Nanjing University, Nanjing 210093, China

[2]Collaborative Innovation Center of Advanced Microstructures, Nanjing University, Nanjing 210093, China

[3]Collaborative Innovation Center of Solid-State Lighting and Energy-Saving Electronics, Nanjing University, Nanjing 210093, China

Correspondence: Jianping Ding (jpding@nju.edu.cn)

## Abstract

The recognition in the 1990s that vortex beams (VBs), paraxial light beams with optical vortices, carry orbital angular momentum (OAM), has benefited applications ranging from optical manipulation to high-dimensional classical and quantum information communications. The transverse profiles of common VBs, e.g., Laguerre-Gaussian beam and high-order Bessel beam, are hollow donuts whose radii grow up with OAM inevitably. The inherently unperfect character of the VBs that the radius is always positively correlated with OAM, restricts the application of the VBs in many scenarios like fiber optic data transmission, spatial OAM mode (de)multiplexing communication, and particle manipulation, which call for VBs to have the same scale with distinct OAM or even the small vortex ring for large OAM. Here, we derived a theory based on the most widely used Laguerre-Gaussian beam to generate a brand new type of VB with OAM-independent radii that moves away from the common unperfect constraint, called Laguerre–Gaussian Perfect Vortex Beam (LGPVB). LGPVBs have the self-similar property like common Laguerre-Gaussian beams but can self-heal after suffering disturbance and always remain 'perfection' when propagating. Our Fourier-space design not only allows us to shape the LGPVB's propagating intensity at will, but it also gives LGPVB the fascinating potential to arbitrarily self-accelerate while still perfectly propagating, self-similar, and self-healing. This customizable self-healing LGPVB, whose properties inform our most expectations of VBs, offers a better alternative for application scenarios of common VBs' in a wide range of areas.

## Introduction

Vortices are common phenomena in nature, from quantum vortices in liquid nitrogen to typhoon vortices and even to spiral galaxies in the Milky Way, manifesting themselves not only in macroscopic matter but also in structured electromagnetic fields. Vortex beams (VBs), the paraxial version of optical vortex fields, carry orbital angular momentum (OAM) proposed by Allen et al.[1] in 1992. VBs are usually characterized by a helical phase front of $\exp(im\phi)$ ($m$ and $\phi$ represent the topological charge (TC) and azimuthal angle, respectively), and possess a OAM of $m\hbar$ per photon ($\hbar$ is the Dirac constant), which can be much greater than a spin angular momentum in the circularly polarized beam since the spin angular momentum is only limited in the range of $[-\hbar, \hbar]$ per photon. Since the values of OAM are theoretically unbounded in VBs, the optical VBs can find multitudinous applications in high-dimensional classical and quantum information communications[2,3], micro-particle manipulation[4], optical measurements[5,6], optical imaging[7,8], and processing[9] and so on.

With a spiral phase, VBs (e.g., Laguerre-Gaussian beam (LGB) and high-order Bessel beam (HOBB)[10]) manifest themselves in a ring shape with a phase singularity at the beams' center[11], and the radius of VBs is always positively correlated with OAM, as shown in Fig. 1. Namely, the radii of VBs carrying greater OAM are certainly larger than those carrying smaller OAM under the same conditions. This property of the OAM-dependent vortex size is deemed as 'unperfect' as it may restrict the application in scenarios like fiber optic data transmission, spatial OAM mode (de)multiplexing communication, and particle manipulation which pose challenges like coupling multiple OAM beams simultaneously into single optical fiber, or special demands for large TCs but small ring diameters. Less mature, but receiving intense research interest, is the attempts to lift this restriction. However, conventional perfect optical vortex fields proposed to date are merely restricted to the two dimensional (2D) transverse plane and cannot maintain perfection when propagating[12,13]. Leaving away from this certain plane, the divergence of conventional perfect optical vortex fields increases rapidly as the OAM enlarges, and thus we refer to these as 2D perfect vortex fields. To the best of our knowledge, there is no report yet on the three dimensional (3D) perfect VBs, which can continue to be perfect even after propagation. Here, to address this key issue, we present a brand new type of VB to solve this inherent restriction of VBs' radii, which holds the propagation-invariant perfectness.

By analyzing the distributions of LGBs, we deduce an analytical expression of the vortex radius and reveal that the vortex radius of LGB is not only positively related to the TC/OAM but also relates to the beam waist and the radial index. To break this limit, we propose and generate the new VB based Laguerre–Gaussian mode, named Laguerre–Gauss Perfect Vortex Beam (LGPVB), whose vortex radius can be arbitrarily specified regardless of the OAM. The proposed new VB will always remain 'perfect' when propagating with OAM-independent divergence and can still self-heal after suffering perturbed or impaired. Furthermore, in application scenarios like underwater optical communication, micromanipulation of particles in solutions, and imaging for biological samples, the intensity of common VBs tends to fade out due to the possible scattering or absorption. Naturally, it becomes appealing to manage the intensity evolution of VBs under propagation, but the relevant report on the intensity manipulation of LGBs is still lacking. Aiming at making the LGPVB more versatile and robust under different circumstances, we develop a Fourier-space theory to regulate at will its intensity evolution and propagation trajectory simultaneously. The proposed scheme is verified by both numerical simulation and optical experiments.

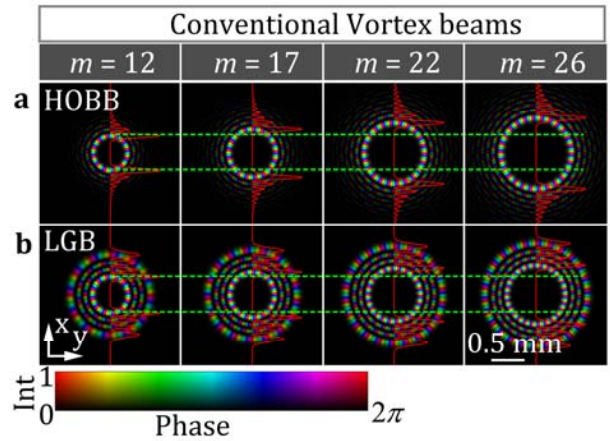

Fig. 1 The complex amplitude distributions of conventional VBs. **a** HOBBs with TCs of 12, 17, 22 and 26, respectively. **b** LGBs with TCs of 12, 17, 22 and 26, respectively, but the same radial index 3 and beam waist 146.5um. The luminance and color of colormap refer to the intensity (Int) and phase of the focal field, respectively. The red curves represent intensity profiles along the x-axis and the horizontal green-dashed lines serve as a reference for the radius of the conventional vortex beams with the TC of 12 in **a** and **b**.

## Results
### Removing the restriction for VBs: LGPVB

The complex amplitude distributions of a LGB in cylindrical coordinates $(r, \varphi, z)$ is represented by

$$U_{m,p}(r,\varphi,z) = \left[\frac{W_0}{W(z)}\right]\left[\frac{r}{W(z)}\right]^m L_p^m\left[\frac{2r^2}{W^2(z)}\right]\exp(-\frac{r^2}{W^2(z)})$$
$$\times \exp\left[ikz + ik\frac{r^2}{2R(z)} + im\varphi - i(m+2p+1)\zeta(z)\right],\quad (1\text{-}1)$$

where $L_p^m$ denotes the Laguerre polynomial with the azimuthal index (or TC) $m$ and the radial index $p$, and $k=2\pi/\lambda$ is the wavenumber with $\lambda$ being the wavelength, and

$$W(z)=W_0\sqrt{1+\left(\frac{z}{z_0}\right)^2},\ R(z)=z[1+\left(\frac{z_0}{z}\right)^2],\ \zeta(z)=\tan^{-1}\left(\frac{z}{z_0}\right),\ z_0=\frac{\pi W_0^2}{\lambda},\ W_0=\sqrt{\frac{\lambda z_0}{\pi}}.\quad (1\text{-}2)$$

with $W_0$ denoting the beam waist. The width of the LGB expands/shrinks upon propagation after/before the beams' waist plane while maintaining their intensity patterns, i.e., having form-invariant property[14]. The LGB's radius varies with the TC, as shown in Fig. 1**b**, and thus it is deemed as unperfect. After deeply analyzing Eq. (1), we get the following analytical solution for the LGB's mainlobe radius (i.e., the vortex ring radius of the LGB) (the detailed derivation can be found in Materials and Methods):

$$r_m(z)=W_0\sqrt{1+\left(\frac{\lambda z}{\pi W_0^2}\right)^2}\frac{C_1\sqrt{(m+1)m}+C_2/2}{\sqrt{2p+m+1}},\quad p\geq 5, m\geq 4,\quad (2)$$

where $r_m(z)$ is the vortex radius of the LGB with the TC $m$ at $z$ position and $C_1$ (=0.5207) and $C_2$ (=0.7730) are constant coefficients. Notably, the vortex radius of the LGB at any $z$ position is not only related to the TC $m$, but to the beam waist $W_0$ and the radial index $p$ as well. This inspires us that we may realize a perfect VB whose radius is OAM-independent by respecifying $p$ and $W_0$ consciously, since the most concern of "perfectness" is that the VB's size does not vary with the TC $m$. However, such a vortex is merely perfect at a specific transverse plane, say, the beam waist plane at $z=0$, and cannot maintain perfectness when propagating away from this plane because different beam waist leads to different divergence, similar to the aforementioned 2D perfect vortex[12,13]. The relevant analysis regarding such a limitation of the 2D perfect vortex can be found in Supplementary Text. To enable the divergence angle of the perfect vortex to be identical for different TC $m$, i.e. OAM-independent, the beam waist $W_0$ in Eq. (2) should remain unchanged so that the desired 3D propagation-perfect vortex with a OAM-independent radius and divergence is achieved.

Aiming at the 3D perfectness, the desired vortex beams must hold a OAM-independent radius at each transverse plane, i.e. having propagation-invariant form up to a scaling. For this purpose, let us focus on a single position, e. g. the beam waist plane ($z=0$), and accordingly the vortex radius at the beam waist plane can be expressed as

$$r_m(0)=W_0\frac{C_1\sqrt{(m+1)m}+C_2/2}{\sqrt{2p+m+1}},\quad p\geq 5, m\geq 4.\quad (3)$$

When $m$ changing, we can choose appropriate radial index $p(m)$ to ensure the fixed radius, $r_m(0)$, for the vortex beam, which thus has OAM-independent width. Consequently, the appropriate radial index $p(m)$ can be expressed as

$$p(m)\approx \left[W_0\frac{C_1\sqrt{(m+1)m}+C_2/2}{4r_m(0)}\right]^2 - \frac{(m+1)}{2},\quad p\geq 5, m\geq 4\quad (4)$$

and the distributions of this brand new perfect vortex beams can be explicitly expressed as

$$U_{m,p(m)}(r,\varphi,z)=\left[\frac{W_0}{W(z)}\right]\left[\frac{r}{W(z)}\right]^m L_{p(m)}^m\left[\frac{2r^2}{W^2(z)}\right]\exp(-\frac{r^2}{W^2(z)})$$
$$\times\exp\left[ikz+ik\frac{r^2}{2R(z)}+im\varphi-i(m+2p(m)+1)\zeta(z)\right]. \quad (5)$$

Obviously, this new perfect vortex beam also belongs to the LGB mode while having the excellent propagation-invariant perfectness, thus being named the Laguerre–Gauss Perfect Vortex Beam. The LGPVB with high-order radial index $p$ can exhibit the extra property of self-healing[15], in the sense that they tend to reconstitute themselves even when they have been severely perturbed or impaired, making themselves more robust (shown in Fig. 2**g** and **h**). Besides, the radial index $p$ must be an integer in the Laguerre polynomial, thus the solution of Eq. (4) should take the nearest integer. Remarkably, the vortex radius is proportional to the beam waist as indicated by Eq. (3).

A common constraint of those conventional vortex beams is that the vortex radius increases with its TC, as shown in Fig. 1. HOBBs in Fig. 1**a**, with different TCs of 12, 17, 22 and 26, respectively, and meanwhile with the same radial wavenumber[16] of $0.0047k_0$, have their respective vortex radii of 0.250 mm, 0.335 mm, 0.439 mm and 0.515 mm; LGBs in Fig. 1**b**, having TCs of 12,17,22 and 26, respectively, and meanwhile with the same radial index of 3 and the beam waist of 0.147 mm, have their respective vortex radii of 0.250 mm, 0.313 mm, 0. 368 mm and 0.406 mm. In stark contrast to Fig. 1, the LGPVBs, shown in Fig. 2**a-c** (simulation) and **d-f** (experiment) generated from Eqs. (4) and (5) with setting $r_m(0)=0.719$ mm, $W_0=0.374$ mm and TC $m$ = 19, 21, and 25, always have the same radius when propagating and thus remove this "imperfectness" of common LGBs. Because the LGB represents the most popular and used laser beam, endowing the ability of arbitrarily designing vortex radius regardless of the TC to LGPVB will bring more opportunities to the application of optical vortices in the field like the next generation of high-capacity OAM optical communication, and may pave the way for the integration and miniaturization of related optical devices.

LGPVBs can exhibit another intriguing and useful self-healing property of tending to reconstitute their shape even when they have been severely perturbed or impaired, as shown in Fig. 2**g** (see also Supplementary Video S1), where the LGPVB (with the parameters of $m$ = 19, $p$ = 12, $W_0$ = 0.159 mm and $z_0$ = 150 mm) encounters a square obstacle at z = -150 mm and then gradually reconstruct themselves after propagating through a distance. Energy circulation can provide an insightful picture of self-healing process, and the momentum density has been calculated, following from the cycle-average Poynting vector[17] given by $\boldsymbol{p}=\varepsilon_0/(2\omega)\text{Im}\left[\boldsymbol{U}^*\times(\nabla\times\boldsymbol{U})\right]$, where $\varepsilon_0$ and $\omega$ are the vacuum permittivity and circular frequency, respectively. Hence we evaluate the transversal momentum density as below. Fig. 2**h** (also Supplementary Video S2) illustrates the calculated transverse orbital flow density of Fig. 2**g** and the red arrows indicate the value and direction of each flow. The evolution of the transverse momentum density reveals the fact that the reconstruction of the damaged intensity pattern is dominated by the azimuthal constituent of momentum density. Interestingly, the radial constituent appears to flow inward from the outer sidelobes, thereby contributing to the beam's resilience robustness against the diffractive spreading. The LGPVB with self-healing property will facilitate its applications under disturbed turbid circumstances like atmospheric turbulence, seawater, and biological tissues.

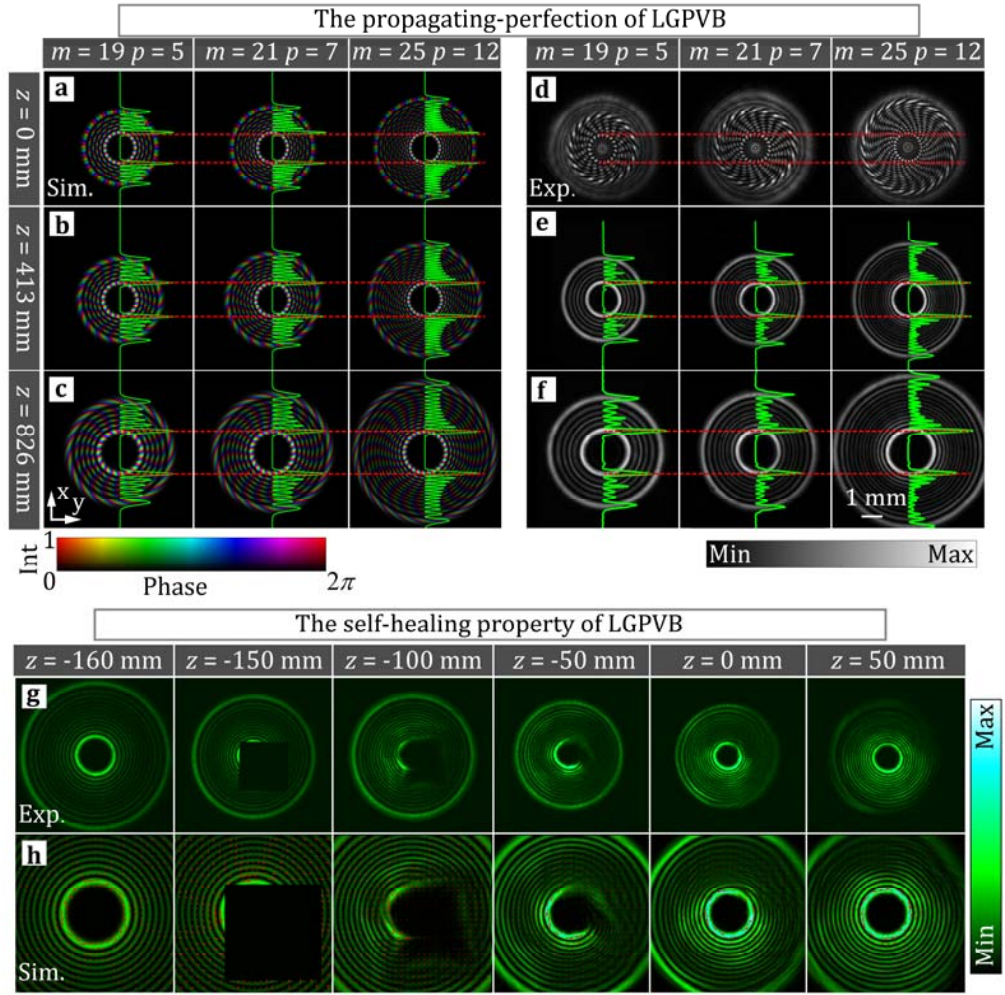

Fig. 2. The perfect and self-healing characters of LGPVBs. The complex amplitude distributions of LGPVBs (having the beam waist 0.374 mm and the Rayleigh range 826 mm but different TC $m$ and radial index $p$) at **a** $z$=0 mm (the beam waist plane), **b** $z$=413 mm, **c** $z$=826 mm, wherein the luminance and color of colormap refer to the intensity (Int) and phase of the focal field, respectively. **d-f** the corresponding experimental results for **a-c** (especially, **d** is the interference patterns between the LGPVBs and a reference plane wave). In **a-f**, the green curves represent intensity profiles along the $x$-axis and the horizontal red-dashed lines serve as a reference for indicating the diameter of the LGPVB's vortex ring. The demonstration of self-healing property when the LGPVB ($m$=19, $p$=12, $W_0$=0.159 mm, $z_0$=150 mm) is blocked by a square obstacle at $z$=-150 mm: **g** the experimental results of transversal intensity maps at different z-axial locations (Supplementary Video S1); **h** the corresponding transversal momentum density of **g** marked by red arrows (Supplementary Video S2);

## Theory for Customizing LGPVB

**Customizing Intensity.** Although the LGPVBs are robust against disturbed turbid circumstances because of their self-healing property, their intensity still tends to fade out due to the possible scattering or absorption in a lossy medium. Naturally, it becomes appealing to control and compensate the intensity of these beams along the propagation direction. To this end, we propose to customize the LGPVB intensity and give the theoretical derivation as follows.

A monochromatic light field propagating along the z-axis can be represented by its angular spectrum in cylindrical coordinates $(r, \varphi, z)$ as

$$U(r,\varphi,z) = (1/2\pi)^2 \iint A(k_r,\phi) e^{ik_r r \cos(\phi-\varphi)} e^{ik_z z} k_r dk_r d\phi \tag{6}$$

where $(k_r, \phi, k_z)$ represent the three-dimensional cylindrical coordinates in the Fourier-space (**k**-space), respectively, and $A(k_r, \phi)$ is the angular spectrum of the light field $U(r, \varphi, z = 0)$ in the focal plane at $z = 0$. The light field $U(r, \varphi, z)$ must obey the Helmholtz Equation by retaining the **k**-space relation $k_z = \sqrt{k^2 - k_r^2}$ while neglecting evanescent waves. The Fourier transformation expressed in Eq. (6) can be optically realized by a lens focusing process (shown in Fig. 3**a**): an incident light field distribution $A(k_r, \phi)$ in the front focal plane of the lens (focal length $f$) will be transformed into the focal field $U(r, \varphi, z=0)$ in the rear focal plane at $z = 0$, then yielding a light field distribution $U(r, \varphi, z)$ in an arbitrary plane at z after accounting for a free-space propagator $\exp(ik_z z)$ through Eq. (6). It should be noted that in the real-space coordinate system the radial wavenumber $k_r$ must be converted into the radial coordinate $\rho$ in the incident plane (i.e. the front focal plane) according to the relation $k_r = k\rho/f$.

The LGPVB expressed in Eq. (5) can be factorized into the product of radial and azimuthal parts as $U(r, \varphi, z) = U(r, z) \exp(im\varphi)$, and so is its angular spectrum at $z = 0$ in the form of $A(k_r, \phi) = A(k_r) \exp(im\phi)$. We get from Eq. (6) the following expression for the field distribution of the focal region by using the Bessel identity and $k_r = \sqrt{k^2 - k_z^2}$,

$$U(r,z) = 2\pi i^m \int_0^k A\left(\sqrt{k^2 - k_z^2}\right) J_m(\sqrt{k^2 - k_z^2} r) e^{ik_z z} k_z dk_z, \tag{7}$$

where $J_m$ is the $m$th-order Bessel function. Eq. (7) implies that the LGPVB with the TC $m$ can be seen as a composition of ideal $m$th-order HOBBs with different complex weighting factors. We merely pay attention to the complex amplitude distribution on the LGPVB's vortex-ring, namely,

$$U_{m,p(m)}(r = r_m(z), z) = Amp(z) \exp\left[-ikz - ik\frac{r_m^2(z)}{2R(z)} + i(m+2p+1)\tan^{-1}\left(\frac{z}{z_0}\right)\right], \tag{8}$$

where $Amp(z)$ is the normalized amplitude on the vortex ring, expressed by

$$Amp(z) = \left[\frac{W_0}{W(z)}\right]\left[\frac{r_m(z)}{W(z)}\right]^m L_{p(m)}^m \left[\frac{2r_m(z)^2}{W^2(z)}\right] \exp(-\frac{r_m(z)^2}{W^2(z)}). \tag{9}$$

The needed incident light field is calculated by the inverse transform of Eq. (7):

$$A\left(\sqrt{k^2 - k_z^2}, \phi\right) = \frac{e^{im\phi} \int_{-\infty}^{\infty} Amp(z) \exp\left[ikz + ik\frac{r_m^2(z)}{2R(z)} - i(m+2p(m)+1)\tan^{-1}\left(\frac{z}{z_0}\right)\right] e^{-ik_z z} dz}{2\pi i^m \text{rect}\left(\frac{k_z}{2k}\right) J_m(\sqrt{k^2 - k_z^2} r_m) k_z}$$

$$= \frac{e^{im\phi} FT\left[Amp(z) \exp\left[ikz + ik\frac{r_m^2(z)}{2R(z)} - i(m+2p(m)+1)\tan^{-1}\left(\frac{z}{z_0}\right)\right]\right]}{2\pi i^m \text{rect}\left(\frac{k_z}{2k}\right) J_m(\sqrt{k^2 - k_z^2} r_m(z)) k_z}, \tag{10}$$

where rect( ) represents the rectangle function and $FT$ denotes the Fourier transform operation. Replacing the amplitude distribution $Amp(z)$ of LGPVBs in Eq. (9) with a specified intensity profile, say $I(z)$, the corresponding incident light field (angular spectrum) can be expressed as

$$A\left(\sqrt{k^2-k_z^2},\phi\right)=\frac{e^{im\phi}\int_{-\infty}^{\infty}\sqrt{I(z)}\exp\left[ikz+ik\frac{r_m^2(z)}{2R(z)}-i(m+2p(m)+1)\tan^{-1}\left(\frac{z}{z_0}\right)\right]e^{-ik_zz}dz}{2\pi i^m \text{rect}\left(\frac{k_z}{2k}\right)J_m(\sqrt{k^2-k_z^2}r_m)k_z}$$

$$=\frac{e^{im\phi}FT\left[\sqrt{I(z)}\exp\left[ikz+ik\frac{r_m^2(z)}{2R(z)}-i(m+2p(m)+1)\tan^{-1}\left(\frac{z}{z_0}\right)\right]\right]}{2\pi i^m \text{rect}\left(\frac{k_z}{2k}\right)J_m(\sqrt{k^2-k_z^2}r_m(z))k_z}.$$

(11)

In this way, the *z*-directional amplitude distribution of LGPVBs can be customized as $\sqrt{I(z)}$. Through Fourier transforming the designed angular spectrum (optically realized by a lens focusing process in Fig. 3**a**), those LGPVBs (or LGBs) with the desired intensity *I*(*z*) will be generated. Besides the spiral phase *mϕ*, there exist two other types of characteristic phase in Eq. (11); the Gouy phase, $-(m+2p(m)+1)\tan^{-1}(z/z_0)$, corresponds to an excess delay of the wavefront and the curvature phase, $k(r_m(z))^2/(2R(z))$, is responsible for wavefront bending. In most scenarios, the curvature phase is relatively slowing varying function comparing with the Gouy phase (the varying rate is at least 1 to 2 orders of magnitude slower). For simplicity, we rewrite Eq. (11) as below by ignoring the curvature phase

$$A\left(\sqrt{k^2-k_z^2},\phi\right)=\frac{e^{im\phi}\int_{-\infty}^{\infty}\sqrt{I(z)}\exp\left[ikz-i(m+2p(m)+1)\tan^{-1}\left(\frac{z}{z_0}\right)\right]e^{-ik_zz}dz}{2\pi i^m \text{rect}\left(\frac{k_z}{2k}\right)J_m(\sqrt{k^2-k_z^2}r_m(z))k_z}$$

$$=\frac{e^{im\phi}FT\left[\sqrt{I(z)}\exp\left[ikz-i(m+2p(m)+1)\tan^{-1}\left(\frac{z}{z_0}\right)\right]\right]}{2\pi i^m \text{rect}\left(\frac{k_z}{2k}\right)J_m(\sqrt{k^2-k_z^2}r_m(z))k_z}.$$

(12)

For verifying the validity of Eq. (12), the numerical simulation and optical experiment on customizing the LGPVB with the designed intensity profile are shown in Fig. 3. The intensity maps in the x-z plane of common LGB and the LGPVBs having three predesigned intensity profiles, uniform one with *I*(*z*) = rect(*z*/2*f*), linearly increased one with *I*(*z*) = 0.5(-*z*/*f*+1), and exponentially increased one with *I*(*z*) = exp(10*z*)/exp(10*f*), are shown in Figs. 3**b**-3**e**. The corresponding experimental results (blue curves) of intensity measured at the vortex ring are shown in Figs. 3**f**-3**i**, which are in good agreement with the simulation results (red curves). Experimental visualizations for Figs. 3**c**-3**e** are shown in Supplementary Videos S3-S5, respectively. The examples of increased intensity profiles in Figs. 3**d** and 3**e** indicate that the sidelobes of the LGPVB serve as a "reservoir" to store energy for the vortex ring; such behavior of the LGPVB is different from the common LGB. For a clear comparison between the LGB and the LGPVB designed according to Eq. (12), we present in Figs. 3**j**-3**l** the radial intensity profiles at three positions of *z* = -50 mm, 0 mm, and 100 mm and in the Fig. 3**m** the vortex radius divergency along the z-direction for the LGB in Fig. 3**b** and the LGPVB in Fig. 3**c**. The comparison manifests that the customized LGPVB from Eq. (12) has the same behavior of vortex width evolution as the LGB. Hence, the LGPVBs with controllable propagating intensity can be seen as the quasi-LG mode beam, which means they inherit the structurally stable characteristics of LGB but can be manipulated more flexibly. Another interesting observation on Figs. 3**b**-3**e** implies that the regulation of the vortex ring (the innermost bight ring) is mediated by the energy redistributing of the outer subsidiary rings of the LGPVBs. This is the first time to realize the manipulation of the propagating property of LG mode beam. Acquiring the ability to shape the propagating intensity profile of LGPVBs can benefit for applications that utilize light beams in turbid absorbing media such as biological tissues or fluids which limit the beam propagation range due to the light intensity decay.

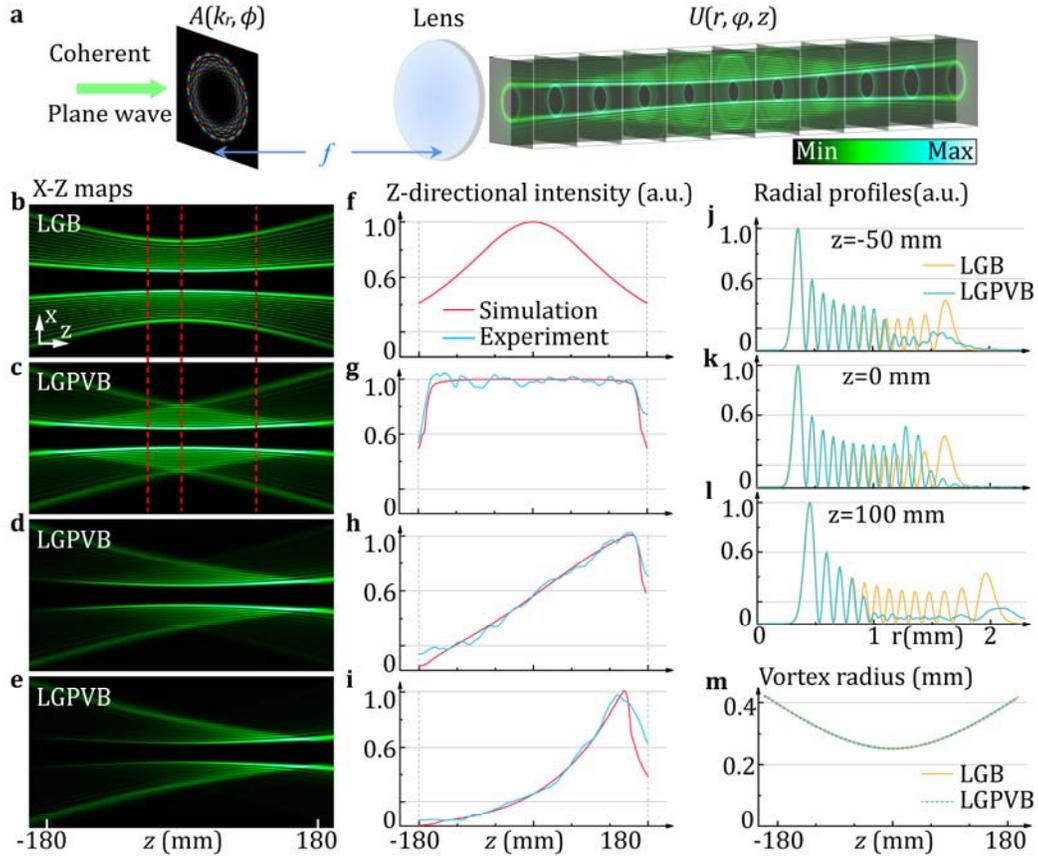

Fig. 3. Customizing the propagating intensity of LGPVB. **a** Schematic of customizing LGPVB in the Fourier space. **b** the intensity map of common LGB in the x-z plane with $m = 19$, $p = 12$, $W_0 = 0.159$ mm, $z_0 = 150$ mm. Intensity maps of LGPVB with same parameters but predesigned **c** uniform $I(z) = \text{rect}(z/2f)$, **d** linearly increased $I(z) = 0.5(-z/f+1)$, and **e** exponentially increased $I(z) = \exp(10z)/\exp(10f)$ intensity profiles along the z-direction. Red dashed lines in **b** and **c** mark positions at $z = -50$ mm, 0 mm, and 100 mm. **f-i** the corresponding z directional intensity profiles of the vortex rings in **b-e**. The red (blue) curves depict the simulation (experimental) results. **j-l** the radial profiles of **a** LGB (brown curves) and **b** LGPVB with uniform intensity distribution (cyan curves) at the position of $z = -50$ mm, 0 mm, and 100 mm respectively. **m** the vortex radius diverges along the z-direction of **a** common LGB (brown curves) and **b** LGPVB with uniform intensity distribution (cyan dot curves). Experimental visualizations for **c-e** are shown respectively in Supplementary Videos S3-S5.

To directly elucidating the practicability of the propagating intensity shaping, we have carried out another experiment to observe the propagation behavior of the LGB and LGPVB in a lossy media. We let the beams transmit through a milk suspension and measure the cross-sectional intensity at different z positions. Fig. 4**a** shows the beam trajectory visualized through to the scattering effect. To characterize the light intensity attenuation in the milk suspension, the measured intensities are fitted with an exponential curve, which turns out to be $\exp(-3.5z)$. Similarly, the transversal intensity maps at different z-axial locations of the traditional LGB and the attenuation-compensated LGPVB with predesigned exponentially increased intensity profile $I(z) = \exp(3.5z)/\exp(3.5f)$ in this lossy medium are recorded and shown in Fig. 4**b** and **c**, respectively. The intensity values, normalized at their maximum, on the vortex ring of the LGB and the attenuation-compensated LGPVB at different z locations have been presented in red font at the bottom of Fig. 4**b** and **c**, respectively, and also plotted in Fig. 4**d** with the blue triangles and blue circles, respectively. Obviously, although the total energy is attenuated, the sidelobes of the attenuation-compensated LGPVB in Fig. 4**c** compensate the energy loss on the vortex ring and the intensity profile on the vortex ring along the z-direction can remain almost uniform. To show the feasibility of attenuation compensation in different media, another milk suspension (i.e., medium 2 in Fig. 4**d**) with an attenuation curve of exp(-7.2$z$) is also tested. The intensity profiles on the vortex ring of the common LGB and the attenuation-compensated LGPVB with predesigned exponentially increased intensity profile $I(z)$ = exp(7.2$z$)/exp(7.2$f$) are measured and plotted in Fig. 4**d** with the red triangles and red circles, respectively. The intensity profile of this attenuation-compensated LGPVB remains almost constant as expected. The attenuation-compensated LGPVBs may benefit applications of structured light fields to complex circumstances, for example, underwater optical communication and deep biological imaging.

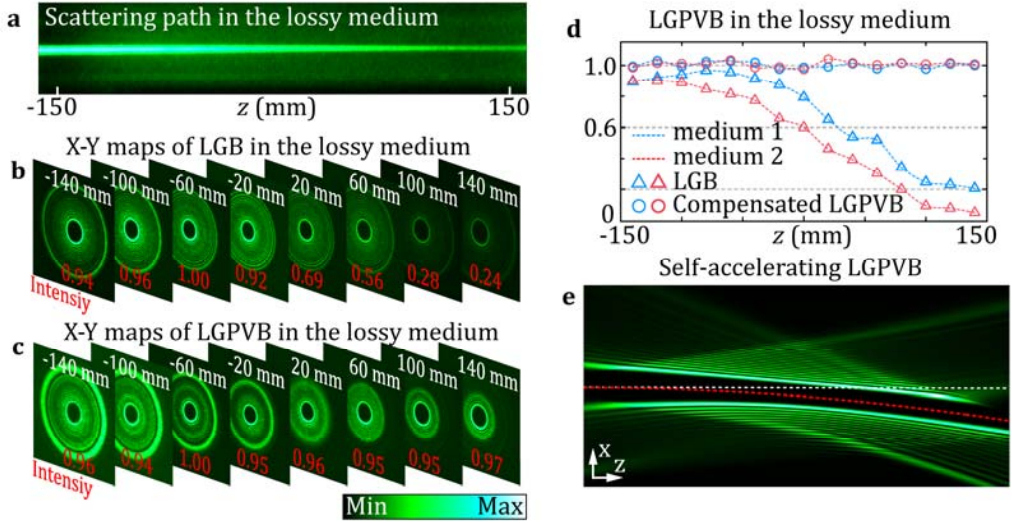

Fig. 4. Experimental demonstration for attenuation-compensated LGPVB in a lossy medium. **a** The scattered path when LGB or customized LGPVB propagates in medium 1 (milk suspension with a measured attenuation curve exp(-3.5$z$)). **b** The cross-sectional intensity maps the common LGB and **c** the attenuation-compensated LGPVB with predesigned exponentially increased $I(z)$ = exp(3.5$z$)/exp(3.5$f$) intensity profiles in medium 1 at $z$ = −140 mm, -100 mm, -60 mm, -20 mm, 20 mm, 60 mm, 100 mm, and 140 mm, respectively. The red numbers in intensity maps represent the measured intensity on the vortex ring which are also plotted in **d** with the blue triangles for the common LGB and blue circles for the attenuation-compensated LGPVB. The red triangles and red circles in **d** represent the measured intensity on the vortex ring of the common LGB and attenuation-compensated NDPVBs with another predesigned exponentially increased profile $I(z)$ = exp(7.2$z$)/exp(7.2$f$) in medium 2 that has the attenuation curve $I(z)$ = exp(-7.2$z$). **e** The self-accelerating parabolic LGPVB with $m$ = 19, $p$ = 12, $W_0$ = 0.159 mm, $z_0$ = 150 mm, $s(z)$ = -1.6×10$^{-4}$ [0, (1+$z$/$f$)$^2$], in which the white dot line serves as a reference for the optical axis and the red dot curve depicts the center of self-accelerating parabolic LGPVB. Experimental visualizations for **e** are shown in Supplementary Video S6.

**Customizing Trajectory.** Besides the ability of intensity control, this Fourier-space-based theory has the potential to realize the manipulation of other propagating properties of LGB such as arbitrarily self-accelerating. The further theoretical derivation for arbitrarily self-accelerating LGPVB is as follows.

After Berry and Balazs theoretically predicted Airy beams in 1979, the self-accelerating beams have stimulated substantial research interest and have found a variety of applications, including particle manipulation and propelling[18], bending surface plasmons and electrons[19], curved plasma generation[20], light-sheet microscopy[21], single-molecule imaging[22], and others. There is little doubt that LGPVBs with arbitrarily self-accelerating capability provide more versatility and can also have significant advantages in many fields. Based on the formalism of calculus, let us elucidate our recipe by revisiting the one-dimensional Fourier integral in Eq. (12) [23]; the integral can be regarded as the infinite sum of many field slices, each of which locates in the tiny z-axial range of $(z, z+\Delta z)$ and is expressed by

$$A_z\left(\sqrt{k^2-k_z^2},\phi\right)=\frac{e^{im\phi}\sqrt{I(z)}\exp\left[ikz-i(m+2p(m)+1)\tan^{-1}\left(\frac{z}{z_0}\right)\right]e^{-ik_z z}}{2\pi i^m \operatorname{rect}\left(\frac{k_z}{2k}\right)J_m(\sqrt{k^2-k_z^2}r_m(z))k_z}. \tag{13}$$

$A_z\left(\sqrt{k^2-k_z^2},\phi\right)$ represents the angular spectrum of a light sheet within $(z, z+\Delta z)$. In this way, we can first sculpture all tiny light sheets and then accumulate them to reconstitute the integration to obtain the customized LGPVB. According to the Fourier phase-shifting theorem, the light sheet within $(z, z+\Delta z)$ will undergo a translation from $(x, y, z)$ to $(x-g(z), y-h(z), z)$ if a complex exponential $\exp(ik_x g(z)+ik_y h(z))$ is imposed on the angular spectrum so that it is explicitly expressed by

$$A_z\left(\sqrt{k^2-k_z^2},\phi\right)=\frac{e^{im\phi}\sqrt{I(z)}\exp\left[ikz-i(m+2p(m)+1)\tan^{-1}\left(\frac{z}{z_0}\right)+ik_x g(z)+ik_y h(z)\right]e^{-ik_z z}}{2\pi i^m \operatorname{rect}\left(\frac{k_z}{2k}\right)J_m(\sqrt{k^2-k_z^2}r_m(z))k_z}. \tag{14}$$

where $(k_x, k_y)$ is the transverse wavenumber in Cartesian coordinates with $k_x = k_r\cos\phi$ and $k_y = k_r\sin\phi$. Eq. (14) can be envisioned as a means for controlling the position of LGPVB in the single transverse plane at $z$. Consequently, the synthetic angular spectrum of the self-accelerating LGPVB can be calculated from the sum of all light sheets by

$$\begin{aligned}A\left(\sqrt{k^2-k_z^2},\phi\right)&=\frac{e^{im\phi}}{2\pi i^m \operatorname{rect}\left(\frac{k_z}{2k}\right)k_z}\int_{-\infty}^{\infty}\frac{\sqrt{I(z)}e^{\left[ikz-i(m+2p(m)+1)\tan^{-1}\left(\frac{z}{z_0}\right)+ik_x g(z)+ik_y h(z)\right]}}{J_m(\sqrt{k^2-k_z^2}r_m(z))}e^{-ik_z z}dz \\ &=\frac{e^{im\phi}}{2\pi i^m \operatorname{rect}\left(\frac{k_z}{2k}\right)k_z}FT\left[\frac{\sqrt{I(z)}e^{\left[ikz-i(m+2p(m)+1)\tan^{-1}\left(\frac{z}{z_0}\right)+ik_x g(z)+ik_y h(z)\right]}}{J_m(\sqrt{k^2-k_z^2}r_m(z))}\right].\end{aligned} \tag{15}$$

And the self-accelerating trajectories can also be expressed in terms of the displacement vector $\mathbf{s}(z)=g(z)\hat{x}+h(z)\hat{y}=(g(z),h(z))$ with $\hat{x}$ and $\hat{y}$ denoting the unit vectors of the $x$ and $y$ directions, respectively. As an exemplary validation, Fig. 4**e** shows a self-accelerating parabolic LGPVB with $m = 19$, $p = 12$, $W_0 = 0.159$ mm, $z_0 = 150$ mm and $\mathbf{s}(z) = -1.6\times10^{-4}\,[0,\,(1+z/f)^2]$. Experimental visualizations for this self-accelerating parabolic LGPVB are shown in Supplementary Video S6. Self-accelerating LGPVB can act as a snowblower conveying particles from one compartment to another[24] or convey information to the obscured invisible receiver[25], and will be utilized in a variety of applications in the areas of laser filamentation, beam focusing, particle manipulation, biomedical imaging, plasmon, and material processing, among others.

As we know, the most remarkable property of the families of LGBs is that their width expands upon propagation from beams' waist while maintaining their intensity patterns up to a scaling, namely form-invariant property. Besides the manipulation of propagating intensity and self-acceleration, our scheme can also utilize to tailor the beam's form by sculpturing the vortex radius of each light sheet of quasi-LGPVB in Eq. (13) respectively and endow extra propagating phase by sculpturing the phase of each light sheet, which are beyond the scope of the current article and will be reported in the future.

## Conclusions

The stubborn and inflexible characteristics of common VBs, e.g. LGB and HOBB whose radii grow up with OAM inevitably, hinder their applications. The proposed LGPVBs represent novel 3D perfect vortex fields, whose vortex radius can be arbitrarily and accurately designed irrespective of the OAM magnitude with OAM-independent divergence, thereby solving this inherent restriction of common VBs and offering a much better alternative for OAM-based application scenarios, e.g. transferring larger OAM to the particles trapped in a confined area in optical tweezers and the OAM-multiplexing-based communication in free space. Besides, this LGPVB can self-heal after suffering from disturbances or impairment, making it suitable for applications in turbid circumstances like atmospheric turbulence, seawater, and biological tissues. We have also investigated the internal flow of the LGPVB and revealed the dynamic origin of its self-healing nature. Furthermore, to compensate for the intensity decay in attenuating media, we have developed a Fourier-space theory to shape the propagating intensity and trajectory of the LGPVB at will. For elucidating the practicability of shaping the propagating intensity profile, we have verified the LGPVBs with customized propagating intensity profiles not only in the air but also in turbid attenuating media. The experimental results prove that it is possible to generate the attenuation-compensating LGPVB with uniform intensity in different media. To conclude, this is the first time, to the best of our knowledge, to realize the manipulation of the propagating properties of LGB, and the malleability of customized LGPVB greatly improves the adaptability of OAM-carrying light beams in different application scenarios.

## Materials and Methods:
### The derivation of Eq. (2): the solution of vortex radius of LGB.

The amplitude distribution of LGB can be written as

$$Amp(r,\varphi,z) = \left[\frac{W_0}{W(z)}\right]\left[\frac{r}{W(z)}\right]^m \exp\left(-\frac{r^2}{W^2(z)}\right)L_p^m\left[\frac{2r^2}{W^2(z)}\right]. \tag{16}$$

Using the approximate relations in Ref.[13,15], the right side of Eq. (16) becomes

$$\left[\frac{r}{W(z)}\right]^m \exp\left(-\frac{r^2}{W^2(z)}\right)L_p^m\left[\frac{2r^2}{W^2(z)}\right] \approx \frac{\Gamma(p+m+1)}{p!(2N)^{m/2}} J_m\left(2\sqrt{2N}\frac{r}{W(z)}\right), p \geq 5, N = p + \frac{m+1}{2}. \tag{17}$$

The radius of LGBs' vortex ring where the maximum amplitude locates can be derived by searching for the following minimum non-zero solution of $\partial A_{mp}(r,\varphi,z)/\partial r = 0$. Accordingly, it follows that

$$\frac{dJ_m\left(2\sqrt{2N}\frac{r}{W(z)}\right)}{dr} = 0. \tag{18}$$

Letting $x = 2\sqrt{2N}r/W(z)$ and using the relation of Bessel function[26], we have

$$\frac{dJ_m(x)}{dx} = \frac{1}{2}(J_{m-1}(x) - J_{m+1}(x)) = 0, \quad x > 0, m > 0. \tag{19}$$

Although Bessel functions are infinitely extending functions with infinite extreme points, we need only to pay attention to the non-zero minimum satisfying Eq. (19) (i.e. the extreme point of the Bessel function closest to the origin). Use the following approximate expression of the Bessel function near the origin[26],

$$J_m(x) \approx \frac{x^m}{2^m m!}, \quad (20)$$

we rewrite Eq. (19) as

$$\frac{x^{m-1}}{2^{m-1}(m-1)!} - \frac{x^{m+1}}{2^{m+1}(m+1)!} = 0, \quad (21)$$

and thereby get the minimum non-zero extreme point of the $m$th-order Bessel function

$$x_m \approx 2m\sqrt{1+\frac{1}{m}}. \quad (22)$$

Owing to the approximation made in Eqs. (17) through (20), the actual position of minimum non-zero extreme point differs from that determined by in Eq. (22), even though by only a tiny percentage. After scrutinizing the approximating function in Eq. (22) and the extremum of Bessel function, we figure out a more accurate relation of $x_m$ versus $m$ by introducing into Eq. (22) two correction constants $C_1$ and $C_2$, namely

$$x_m \approx 2C_1 m\sqrt{1+\frac{1}{m}} + C_2. \quad (23)$$

with $C_1$ and $C_2$ taking the following values:

$$C_1 = \begin{cases} 0.5207, & m \leq 40 \\ 0.5058, & m > 40 \end{cases} \text{ and } C_2 = \begin{cases} 0.7730, & m \leq 40 \\ 2.0223, & m > 40 \end{cases}. \quad (24)$$

When $m>4$, the error of Eq. (23) turns out to be less than 1% while $m \geq 20$, the error is always less than 0.5% and remains less than 0.1% with the larger $m$. Accounting for $x = 2\sqrt{2N}r/W(z)$ we obtain the vortex radii of LGBs at $z$ as below

$$r_m(z) \approx W(z)\frac{C_1\sqrt{(m+1)m} + C_2/2}{\sqrt{2p+m+1}}, \quad p \geq 5, m \geq 4 \quad (25)$$

and Eq. (26) follows immediately at the beam waist plane ($z=0$),

$$r_m(0) \approx W_0 \frac{C_1\sqrt{(m+1)m} + C_2/2}{\sqrt{2p+m+1}}, \quad p \geq 5, m \geq 4. \quad (26)$$

The percent error of Eq. (26) is shown in Fig. 5. When $p<5$, the error appears to be large and negative (blue area in Fig. 5), which is ascribed to the approximation in Eq. (17). When $m<4$, the error is large and positive (red area in Fig. 5), due to the approximation in Eq. (20). Noticeably when $p \geq 5$ and $m \geq 4$, the error remains to be below 5% and becomes even smaller and insignificant as $p$ and $m$ increase. For the width of the vortex ring, this error is negligible in most cases. Noteworthily, this error does not change with the propagation position $z$ and the beam waist $W_0$, and the error map of Eq. (25) is the same as Eq. (26) in Fig. 5.

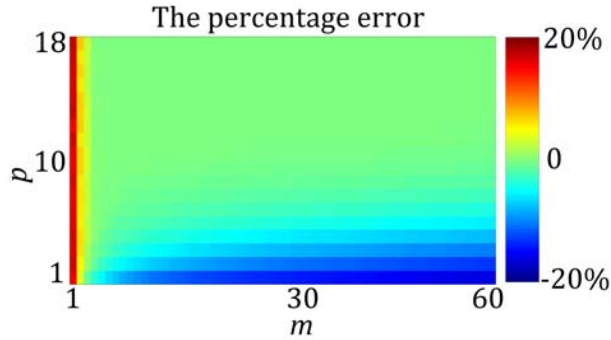

Fig. 5. The percent error of Eq. (26) varying with *m* and *p*. The percentage error is defined as the ratio between the difference of two vortex radii, respectively calculated from Eq. (26) and exact LGBs expression, and the exact LGB's vortex radius.

**Experimental Method**

The experiment setup is shown in Fig. 6. A reflective SLM (Holoeye GAEA-2, 3.7 um pixel pitch, 3840×2160) imprinted with computer-generated hologram patterns (the Fourier marks) transforms a collimated laser light wave at a wavelength of 532 nm into the complex field corresponding to the angular spectrum of LGPVB in the real-space coordinate system, with help of spatial filtering via a 4-F system consisting of lenses L1 and L2, and an iris as well. The resulting field is responsible for generating LGPVBs in the focal volume of lens L3 with a focal length of 200 mm. A delay line, consisting of right-angle and hollow-roof prism mirrors and a translation stage, enables the different cross-sections of LGPVB to be imaged on a complementary metal oxide semiconductor (CMOS) camera (Dhyana 400BSI, 6.5 um pixel pitch, 2048×2040) after a relay 4-F system consisting of two lenses (L4 and L5, each with a focal length of 200 mm). The combination of the delay line and the relay system enables us to record intensity cross-sections at different z-axial locations before and after the focal plane of the lens L3.

The high-order Bessel function term $J_m(\sqrt{k^2 - k_z^2} r_m)$ appears in the denominator in Eqs. (12-15) whose zero points will make the maximum of calculated angular spectrum too large to be correctly loaded on the spatial light modulator. Our numerical simulation and optical experiment show that replacing the high-order Bessel function term in Eqs. (12-15) with 1 can even yield fairly satisfactory LGPVBs. We speculate that the high-order Bessel function is merely too slightly to influence the radial amplitude profile of the ring-shaped angular spectrum and thus can be neglected with inappreciable errors.

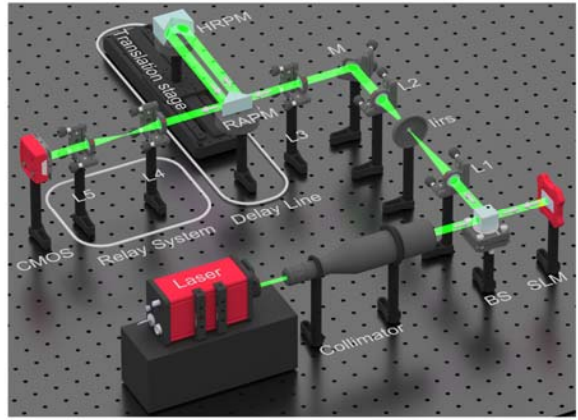

Fig. 6. Experimental setup for generating and detecting LGPVBs. BS, beam splitter; SLM, phase-only spatial light modulator; L1–L5, lens; M, mirror; RAPM, right-angle prism mirror; HRPM, hollow roof prism mirror; CMOS, complementary metal oxide semiconductor camera.

## Acknowledgements

The authors acknowledgment the finance support from the National Key R&D Program of China (Grants Nos. 2018YFA0306200 and 2017YFA0303700), the National Natural Science Foundation of China (Grants Nos. 91750202 and 11922406).

Supplementary information accompanies the manuscript on the Light: Science & Applications website (http://www.nature.com/lsa)

# Supplementary information for

## Customizable Laguerre–Gauss Perfect Vortex Beams

Wenxiang Yan,[1] Zheng Yuan,[1] Yuan Gao,[1] Zhi-Cheng Ren,[1,2] Xi-Lin Wang,[1,2] Jianping Ding,[1,2,3] and Hui-Tian Wang,[1,2]

[1]National Laboratory of Solid State Microstructures and School of Physics, Nanjing University, Nanjing 210093, China

[2]Collaborative Innovation Center of Advanced Microstructures, Nanjing University, Nanjing 210093, China

[3]Collaborative Innovation Center of Solid-State Lighting and Energy-Saving Electronics, Nanjing University, Nanjing 210093, China

Correspondence to: jpding@nju.edu.cn

**This PDF file includes:**

Supplementary Text: Limitation of 2D perfect vortex fields based on Eq. (2).
Supplementary Videos S1-S6.

## Supplementary Text: Limitation of 2D perfect vortex fields based on Eq. (2).

Even though perfect vortex fields can be generated by retuning the radial order $p$ and the beam waist $W_0$ consciously in Eq. (2) by changing $m$, they are merely restricted to the specific transverse plane at $z = 0$ and cannot maintain perfection when propagating with a distinct divergence for the different $m$. We will elaborate upon this limitation as below.

We first consider keeping the radial order $p$ unchanged and only adjusting the beam waist $W_0$ for different $m$ in order to generate perfect vortex fields. We rewrite the relation in Eq. (3) as

$$W_0 \approx r_m(0) \frac{\sqrt{2p+m+1}}{C_1 \sqrt{(m+1)m} + C_2/2}, \quad p \geq 5,\ m \geq 4,\ C_1 = 0.5207,\ C_2 = 0.7730. \tag{S1}$$

Setting the desired radius $r_m(0)$ and radial order $p$ for a given $m$ will yield the corresponding beam waist according to Eq. (S1). For example, if we chose $r_m(0) = 0.722$ mm and $p = 7$ for $m = 16, 21$ and $25$, the corresponding beam waists equal to 0.449 mm, 0.374 mm and 0.333 mm, respectively. The resulting perfect vortex fields at $z = 0$ are shown in Fig. S1.

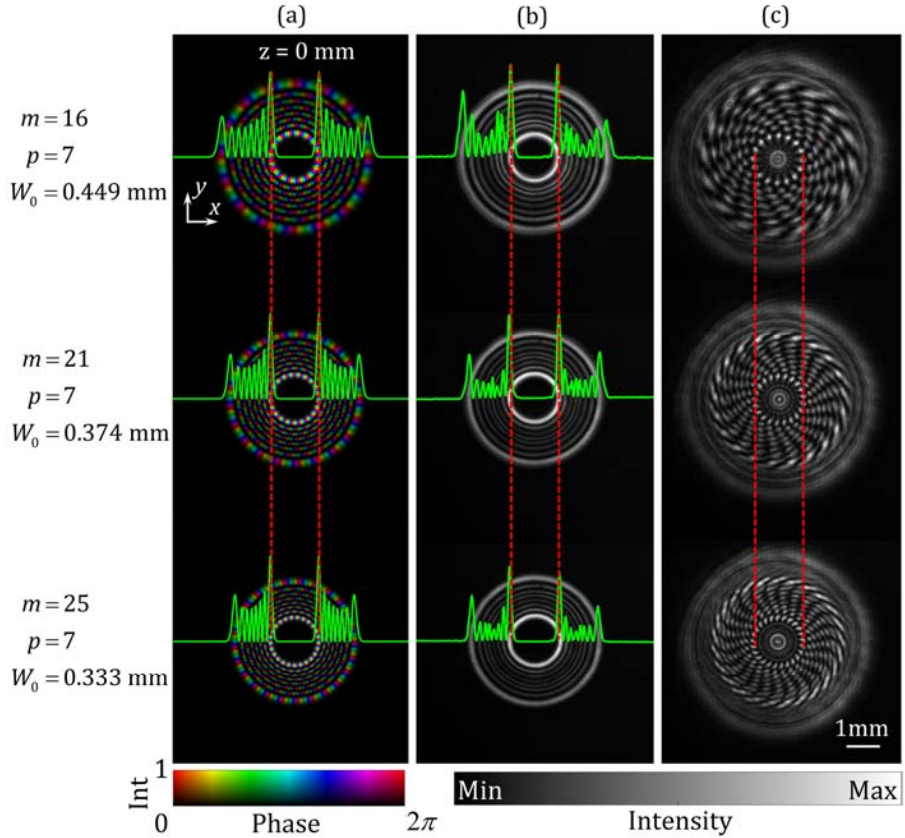

Fig. S1. Generation of perfect vortex fields by regulating beam waists $W_0$ with constant radial order $p$ and different TC $m$. Simulation, experimental, and interference results are shown in column (a-c) respectively. The green curves represent intensity profiles along the $x$-axis and the vertical red-dashed lines serve as a reference for indicating the radius of the perfect vortex field.

Furthermore, for increasing the degree of design freedom, all three parameters $m$, $p$, and $W_0$ are regulated simultaneously in Eq. (3) with the same $r_m(0) = 0.722$ mm, resulting in the perfect vortex fields shown in Fig. S2.

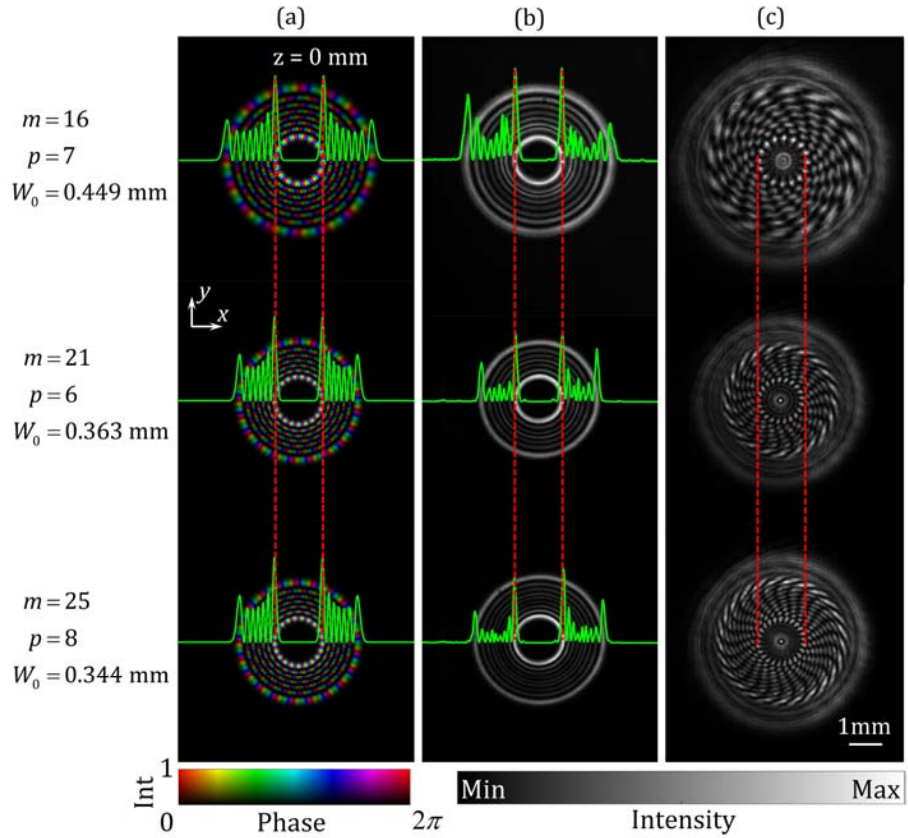

Fig. S2. Generation of perfect vortex fields by regulating all three parameters $m$, $p$, $W_0$. Simulation, experimental, and interference results are shown in column (a-c) respectively. The green curves represent intensity profiles along the x-axis and the vertical red-dashed lines serve as a reference for indicating the radius of the perfect vortex field.

The vortex fields in Fig. S1 and Fig. S2 are actually perfect only in a preset plane (i.e. $z$ = 0) and cannot maintain perfectness when propagating away from this plane. This unperfect behavior can be clearly seen from the distribution in different z planes in Fig. S3. The cause of this imperfection is that the beam widths $W(z)$ have different varying rates along the z-direction due to different divergence. Hence, for generating the propagation-perfect LGPVBs, the beam waist $W_0$ should remain constant for different $m$ to guarantee that $W(z)$ is the same for each LGPVB, i.e. the OAM-independent divergence.

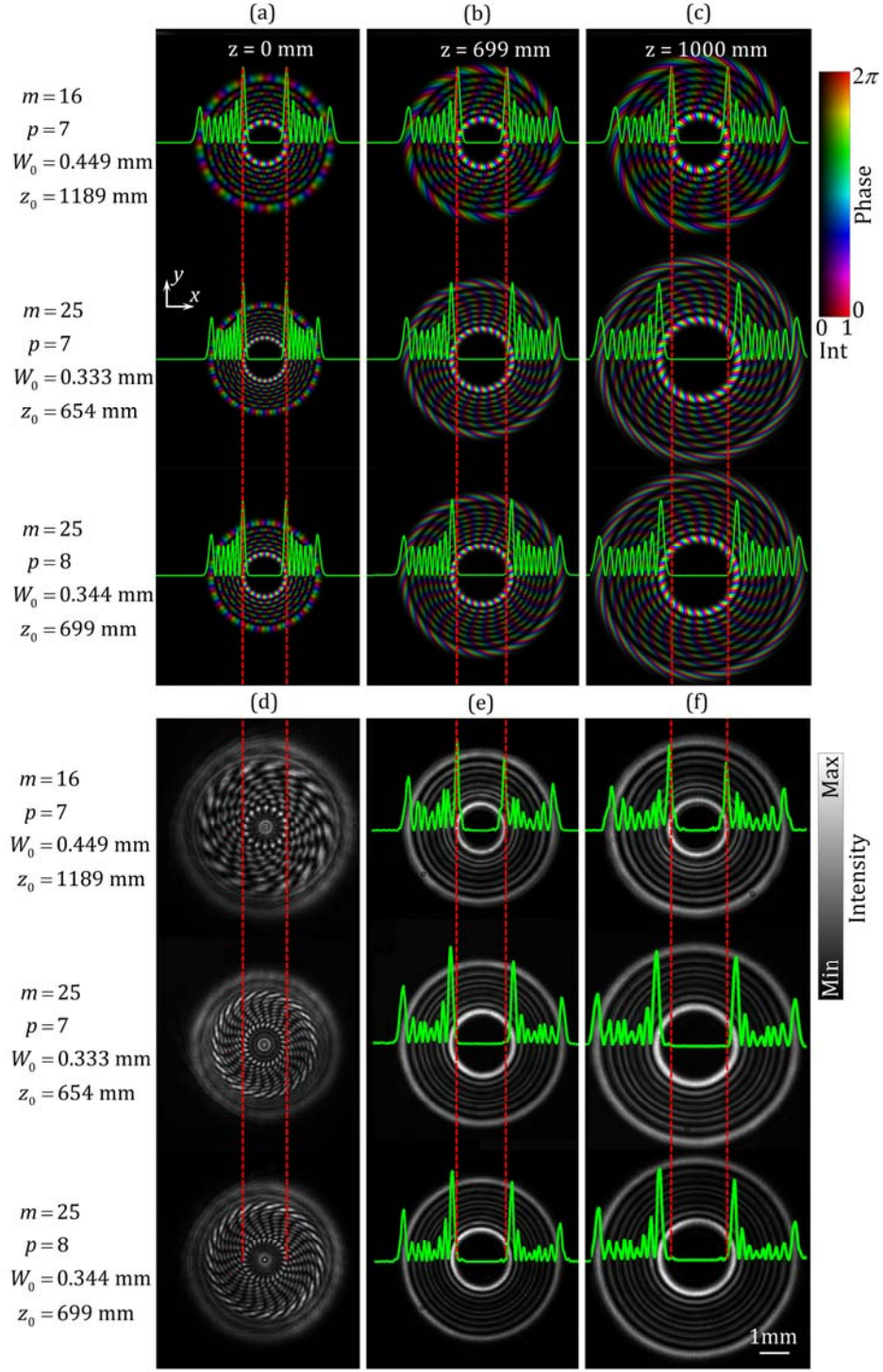

Fig. S3. The 'unperfect' propagation of vortex fields generated in Fig. S1 and Fig. S2 -- their vortex radii are distinctly different at positions away from z = 0. The first to third columns correspond to the distribution of these vortex fields at different z position. (a-c) is the simulation results and (d-f) is the corresponding experimental or interference results. The green curves represent intensity profiles along the x-axis and the vertical red-dashed lines serve as a reference for the radius of the vortex field with the topological charge 16 at different z position.

## Supplementary Videos

Video S1. Experimental movie for the self-healing process of LGPVB in Fig. 2**g**.

Video S2. Corresponding transversal energy flow density of the self-healing process in Fig. 2**h**.

Video S3. Experimental movie for customized LGPVB with uniform longitudinal intensity profile $I(z) = \text{rect}(z/2f)$ with $f$ = 200 mm in Fig. 3**c**.

Video S4. Experimental movie for customized LGPVB with linearly increased longitudinal intensity profile $I(z) = 0.5(-z/f+1)$ with $f$ = 200 mm in Fig. 3**d**.

Video S5. Experimental movie for customized LGPVBs with exponentially increased longitudinal intensity profile $I(z) = \exp(10z)/\exp(10f)$ with $f$ = 200 mm in Fig. 3**e**.

Video S6. Experimental movie for the self-accelerating (self-bending) LGPVB in Fig. 4**e**.